\documentstyle[aps,prl,psfig,epsfig]{revtex}
\tightenlines
\begin{document}
\draft

\title{Mediation of Long Range Charge Transfer by Kondo
Bound States}

\author{R.G. Endres, D.L. Cox, R.R.P. Singh, S.K. Pati\cite{swapan}}               
\address{
Department of Physics, University of California, Davis, CA 95616}
\twocolumn[\hsize\textwidth\columnwidth\hsize\csname
@twocolumnfalse\endcsname

\date{\today}

\maketitle

\begin{abstract}
We present a theory of
non-equilibrium long range charge transfer between donor and
acceptor centers in a model polymer mediated by magnetic exciton (Kondo) bound states.
Our model produces electron tunneling lengths easily exceeding 10$\AA$,
as observed recently in DNA and organic charge transfer systems.
This long ranged tunneling is effective for weak to intermediate
donor-bridge coupling, and is enhanced both by
weak to intermediate strength Coulomb hole-electron attraction (through
the orthogonality catastrophe) and
by coupling to local vibrational modes.
\end{abstract}

]

\narrowtext

Charge transfer (CT) via quantum mechanical tunneling of electrons or
holes
through a molecular bridge between well separated donor (D) and acceptor (A)
sites, is central to many biological processes 
\cite{marcsutin,graydutt,barton,barzewail}, and may play a key role in
molecular electronics applications\cite{ratnernature,newtonscience}.
For proteins, the CT rate typically decays exponentially in the D/A
separation with 
the inverse tunneling length
$\beta = 1.1\pm 0.2\AA^{-1}$\cite{graydutt}.
However, evidence for anomalously
small $\beta$ values ( of order $0.1\AA^{-1}$) has accrued
for CT between metallic intercalating
complexes or organic complexes in DNA\cite{barton,barzewail},
CT from ferrocene attached to  conjugated
polymers\cite{ratnernature,newtonscience}, and a few other
systems \cite{stein}. Values of $\beta\sim 1\AA^{-1}$ are
well accounted for in effective one-electron theories with the D/A levels
typically residing in the middle of the large insulating (4-6 eV) gaps of
the polypeptide bridges of proteins\cite{beratan}. 
Understanding smaller $\beta$ values has generally required ``fine tuning"
of the D/A  levels close to the lowest unoccupied
molecular level (LUMO)\cite{feltsLRCT}. For DNA, this issue
is coupled to the magnitude
of the electronic energy gap in DNA,  
an exciting and hotly debated subject 
\cite{porathDNA,grunerDNA,finkDNA,superconDNA}.

In this paper, we show that long range CT, with tunneling lenghths
$\beta^{-1}\ge 10\AA$ can be achieved 
in model systems through mediation by a semiconducting,
molecular analogue of the Kondo state in which a localized electron or
hole moment is antiferromagnetically coupled to a hole or electron
donated to the bridge\cite{hewson}. While our model for
the tunneling bridge is a conjugated polymer, the conclusions are
generic for a gapped molecule.  The
bound state mechanism provides a robust basis for ``fine tuning'' D
levels close to the LUMO.  We carry out our calculations both with a
semi-analytic variational method\cite{varmayaf} as well as the density
matrix renormalization group (DMRG) approach, and the results agree very
well.
Adding the dielectrically screened electron-hole Coulomb
interaction to the Kondo coupling leads to an enhancement of tunneling for weak to
intermediate Coulomb strength through the orthogonality catastrophe, and
a crossover to deep Coulomb binding for larger Coulomb interaction.
Coupling to local vibrational modes of the metal complexes also enhances
the CT rate, and can lead to normal or inverted regimes
depending upon the strength of the binding and the D/A distance.
 
CT via a single tunneling event
is intiated out of equilibrium either by the
absorption of light (photolysis) or transfer of charge by radicals
(radiolysis) so that an electron or hole at the D complex is energetically
excited above
an empty level in the A complex; in this paper we focus on
photolytically prepared charge transfer. 
In the simplest ``diabatic'' theory for which the
electron tunneling matrix element $H_{DA}$
is small, the CT rate
$k_{ct}$ is well described by a golden rule/linear response approach (Marcus theory\cite{marcsutin}), with
\begin{equation}
k_{ct}(T) = {2\pi \over \hbar} |H_{DA}|^2 F(T)
\end{equation}
where $F(T)$ is the Franck-Condon factor \cite{marcsutin}.
$H_{DA}$, which is the main
subject of this paper, carries the dominant distance dependence.

The CT scenario involves three steps, and can only arise
when, as in Fig. 1,  the excited D complex energy lies at or near the bridge LUMO (a similar
situation can be realized for hole transfer with suitably reversed
energetics).  Upon excitation (step 1), the D complex donates an electron to the
bridge, leaving behind a remnant hole spin with antiferromagnetic
coupling to the bridge electron (this interaction drives the Kondo
binding in step 2).  In step 3, the electron is transferred to the
distant donor level.    
  
Experimental realization of the appropriate D energetics seems
possible. For example,
for the Ru-complex metallointercalator used in DNA CT
experiments\cite{barton}, a 480 nm photoexcitation transfers electrons
from Ru(II) to the ligand ($L^*$) from which electron transfer to the
DNA presumably occurs (there is no direct Ru-DNA overlap), 
leaving a hole spin in the $t_{2g}$ Ru shell.
For DNA gaps of 1-2 eV,
it is plausible that the excited Ru(III)-$L^*$ complex lies
above the electron addition threshold, though estimates of the gap range from
0 eV\cite{finkDNA,superconDNA}, 1-2 eV\cite{porathDNA,sankeyDNA},
and on up to 7 eV\cite{hugeDNA}.
The 3.9 eV photoexcitation energy of the adenine isomer 2-aminopurine (Ap)
is very comparable or in excess of the HOMO-LUMO gap for isolated bases\cite{barzewail},
although the initial CT matrix element to neighboring
DNA bases is likely weaker than in the metallointercalator case, for which
the ligand resides between adjacent base pairs with 
more effective $\pi$-overlap. Excitation of Rh(phi)$_2$(bpy)$^{3+}$
will produce conditions favorable for hole transfer to adjacent
sequences of DNA-A bases\cite{bartonrh,jortner}. For the
5-(p-tolyl)-tetracene donor of ref.\cite{ratnernature}, the D
photoexcitation energy is
nearly equal to the $p$-phenylenevinylene bridge gap for bridges
of 3-5 units, and judging from the trend, should
become higher for longer bridges. Thus, our considerations are clearly relevant
to these systems.
 
With these energetics, the Kondo singlet bound state (KS)\cite{dolg} 
is the most stable
charge zero singlet state, though the charge $\pm 1$ spin 1/2 state (free moment) 
can be globally stable in a molecule
or semiconductor\cite{yu}. Rapid KS formation is required to
compete with other processes, as discussed in Fig. 1.
A lower bound estimate for KS formation is
$\tau_K \approx \hbar/E_B$,
where $E_B$ is the binding energy, in analogy to the reasoning used for
Kondo resonance formation time in quantum dots\cite{nordlander}. $\tau_K$ is
of the order of a few femtoseconds for $E_B=0.1-0.2$eV.  In the
semiconducting/molecular context, it will be divided by a
Franck-Condon factor $F_K$ since the excited D level resides above the
bound state.  Provided $F_K\le 0.001$\cite{triplet},
KS formation will likely
beat luminescence (nanosecond time scale)
and phonon radiative decay to the A state (large polaron formation)
which can be estimated, utilizing polyacetylene parameters\cite{ssh}, to be of order
100 ps.  We stress that (i) no simple polaron dynamics will cause
$k_{ct}$ to decay exponentially in the D/A separation, 
and (ii) any polaron mediated
contribution to $k_{ct}$ is bounded above by the large polaron result\cite{mahan}.

We simply model  
the insulating bridge as a dimerized
tight binding chain of length $N$ ($N$ even) with Hamiltonian ${\cal H}_B$ given by\cite{ssh}
\begin{equation}
{\cal H}_B = - \sum_{i=1,\sigma}^N                                       
(t+(-1)^{i}t')(c^{\dagger}_{i+1\sigma}c_{i\sigma} + h.c.)
\end{equation}
where $c^{\dagger}_{i+1\sigma}$ creates an electron of spin $\sigma$ at
site $i$.  This model describes the $\pi_z$ bands of
polyacetylene, and can easily be extended to other conjugated polymers
(as in refs.\cite{ratnernature,newtonscience}), but the bound state
formation is not special to this model bridge.
For large $N$, this model produces two bands with a full
width of $4t$ and a gap of $2\Delta=4t'$. The
acceptor complex is placed adjacent to site $N$ of the chain, and
modeled as an Anderson
impurity in the empty orbital regime\cite{hewson},
with energy $\epsilon_A$.
To simplify the analysis, we have taken the Coulomb repulsion $U$ between D
holes to be infinite, and as in the Ru case we assume there is a single
hole in an otherwise filled shell, with energy relative to the filled
shell of $\epsilon_D$.
The hybridization Hamiltonian coupling    
the D/A complexes
and bridge is
\begin{eqnarray}
{\cal H}_{BDA} = \sum_{\sigma} [ V_D X_{0\sigma}c_{1,-\sigma} 
+ V_Aa^{\dagger}_{\sigma}c_{N\sigma} + h.c. ] ~~,
\end{eqnarray}
where $|D0>$ is the filled shell state of the Donor
and $|D\sigma>$ the state of a single
hole and $X_{0\sigma}=|D0><D\sigma|$.  We choose $\epsilon_D$ above the
LUMO, and $\epsilon_A$ to be within the gap.  
 
We compute $H_{DA}$ by first identifying the
appropriate two state system from diagonalizing with either:
(1) the DMRG approach, which is exact to within
targeted numerical precision\cite{white}, and (2) a version of the
variational wave function approach for magnetic impurities in metals
pioneered by Varma and Yafet as well as Gunnarsson and                        
Sch\"{o}nhammer\cite{varmayaf}.
We build our variational states by first diagonalizing the
bridge with the non-interacting A ``impurity'' included, and then employ
the {\it Ansatz}:
\begin{equation}
|\psi> = A [ |D0> + \sum_{j\sigma}
\alpha_j b^{\dagger}_{j\sigma} |D-\sigma> 
]|BA>
\end{equation}
where $b_{j\sigma}$ creates an electron in orbital $j$ of the bridge-A
system, $\alpha_j$ are variational coefficients,  
and the particle-hole excitations are controlled both by the
spin degeneracy $N_s$ (here $N_s=2$) and the bridge gap.   As is well
known\cite{hewson}, each added particle-hole excitation
contributes a term of relative order $1/N_s$ to the energy.  This
variational {\it ansatz} does an excellent job of describing the Kondo
state in metals (the second term captures the screening of the
local moment, and
as such is nonperturbative as opposed to, {\it e.g.},
direct  expansion in $U$ or M\"{o}ller-Plesset methods).   Because of the Hilbert-Space truncation associated with
large $U$, the above wave function cannot be reduced to a single
determinant by a unitary transformation.
Hence, for our model, DMRG gives an ``exact configuration
interaction'' solution  while the variational approach provides a
physically motivated basis set reduction solution.   For the purposes of
calculating $H_{DA}$, the methods agree exceedingly well.

Given the two lowest ``adiabiatic'' eigenstates,
we next identify the out-of-equilibrium ``diabatic'' states   
corresponding to the excited electron
predominantly localized to the left (D) side, or right (A) side.  The
Generalized Mulliken-Hush method\cite{cavenewtMH} accomplishes this
by finding, within the two-state space,
the eigenstates for the electric dipole operator,
which maximally localize the charge to left or right. 
$H_{DA}$ is then approximated by the off-diagonal matrix element of the
two-state Hamiltonian in the dipole-diagonal basis. 
  
Our results for $H_{DA}$ vs. bridge length $N$
are shown in Fig. 2(a), where we have set $V_D=V_A$ for
convenience, and chosen $\epsilon_D$ well above the LUMO, and $V_D$ is
of order $t$\cite{vdnote}.  We note the
following features of our curves: (i) for short distances, $H_{DA}$
scales with $V_D V_A$, perturbatively, so that stronger initial coupling
yields stronger tunneling.  On the other hand, for large distances, the
weaker $V_D$ gives better tunneling, because the more loosely bound
KS cloud has better overlap on the distant A complex. 
(ii) By taking the bridge spacing to be of the
order of 3$\AA$ as in DNA, the slope for our curve with $V_D=0.75 t$ at
long distances corresponds to $\beta \approx 0.1\AA^{-1}$, and this
corresponds to a binding energy of the order of 0.05-0.2 eV (see
Fig. 3).  Obviously
our model is too simple to realistically describe CT in DNA; rather, we
stess the ready achievment of small $\beta$ values. 
(iii) The agreement between the DMRG
calculations (x's) and variational (lines) is quantitatively excellent
in the long distance regime where $H_{DA}$ is quite small, and good in
the short distance regime.  Hence, our variational approach is a very
useful, accurate, physically motivated basis set reduction scheme for
this model.  (iv) By varying $t'$, $V_D$, $\tilde\epsilon_D$ we can
easily obtain the typical protein results ($\beta\simeq 1\AA^{-1}$) in
our model. 

Analytic  solutions are possible in two limits.
First, in a continuum limit in which the D/A complexes are placed at
separation $Na$ in a one-dimensional wire, and $E_B<<t'$, we find that
$H_{DA}\sim E_B^{1/4}\exp(-\sqrt{2m^*E_B}Na/\hbar)$, with 
effective mass $m^*=\hbar^2
t'/[2(t^2-t'^2)a^2]$ 
and the Kondo binding energy $E_B =
V^4_D/(\epsilon_0|\epsilon_D-2t'|^2)$, with $\epsilon_0=\hbar^2/(8m^*
a^2)$.   The prefactor
reflects a ``slave boson'' renormalization of the matrix element
$V_D$\cite{hewsonsb}, and the exponent arises from A/KS
overlap.   Thus there is a maximum $H_{DA}$ for
a given bridge length, though this is practically relevant for only
long bridges given the weak prefactor dependence upon $E_B$.
Second, for a                                     
discrete bridge $H_{DA}$ is expressed approximately as
\begin{equation}
|H_{DA}| \approx |V_D V_A\sum_k {\psi(k,1)\psi^*(k,N)\over \epsilon_k
-2t' + E_B}|
\end{equation}
where $\psi(k,i)$ is the wave function amplitude at site $i$
of a bridge-only conduction orbital of energy $\epsilon_k$ (no A
complex).  This describes the propagation from the D complex
to bridge site $1$, then  to bridge site $N$ and onto the A complex
(hole transfer processes are also possible but less dominant).

Given the semiconducting character of the bridge, the dielectrically
screened Coulomb interaction between the D hole and the bridge
electron is of manifest importance.  This is modeled by the interaction
\begin{equation}
{\cal H}_C = -\sum_{j=1}^{N+1} {e^2\over \epsilon ja}[\sum_{\sigma}
X_{\sigma\sigma}] n_j
\end{equation}
where $X_{\sigma\sigma}=|D\sigma><D\sigma|$, and $n_i$ is 
the electron occupancy at site $i$, and $\epsilon$ is the high frequency
dielectric constant.  There are two important effects: (i) For
sufficiently small interaction ($V_c=e^2/(\epsilon a)\le 1$ eV in practice), 
the modification of the bridge wave functions leads to a multiplication
of $V_D$ by $<BA|\bar{BA}>$ where
$|\bar{BA}>$ is the filled valence sea in the presence of the D hole.
This is directly analogous to the orthogonality catastrophe of the core
level x-ray absorption problem.  The effects of this on $H_{DA}$ are shown 
in Fig. 2(b).  Effectively, small $V_c$ reduces $V_D$ and thus enhances
$H_{DA}$. (ii) For larger $V_c$, Coulomb binding overtakes Kondo
coupling, and the stronger binding reduces $H_{DA}$. 

Finally, we introduce coupling to local vibrational
modes of the D/A complexes. We make the harmonic approximation for the 
phonons, and treat the displacements classically, adding to our model  
\begin{eqnarray}
\hat H_{el,ph}\ = \lambda \sum_{\sigma}(x_D X_{\sigma\sigma} + x_A a^{\dagger}_{\sigma}a_{\sigma}) 
\end{eqnarray}
where \(\lambda\) is the electron-phonon coupling constant, 
\(x_D\) and \(x_A\) are the nuclear
coordinates, and we add a harmonic potential $\frac{K}{2}\left(x_D^2\ +\ x_A^2\right)\quad$. 
After the substitution \(x_{D/A}\rightarrow\lambda/K\tilde x_{D/A}\),
with dimensionless coordinates \(\tilde x_{D/A}\), the Hamiltonian depends 
only on a single parameter \(\lambda^2/K\). By neglecting electron-phonon coupling 
within the bridge\cite{assume}, we have calculated the total electronic energy of the 
groundstate and first excited state (two state model) as a function of the nuclear coordinates. 
These constitute the Born-Oppenheimer potential energy surfaces (PES) for the
nuclear motion of the two Anderson impurities. The resonance splitting at
the transition state is $2 H_{DA}$\cite{marcsutin}. 
    
In Fig. 3, we compare with the electron-only Mulliken-Hush approach. Like the Coulomb
interaction discussed previously, coupling to phonons leads to an enhanced 
CT rate. The Mulliken-Hush result corresponds to the limit of vanishing 
phonon coupling. In the perturbative regime \(\lambda^2/K<\!\!<1\), 
we find a linear dependence of \(\ln(H_{DA}/t)\) on \(\lambda^2/K\). 
We can roughly explain this by replacing $\epsilon_D$ in the
asymptotic, electron-only problem with renormalized energy level
\(\tilde\epsilon_D=\epsilon_D+\lambda^2/K\tilde x_D\) and expanding
\(\ln H_{DA}\) to linear order in \(\lambda^2/K\). This result
varies little with 
\(\tilde x_D\), since
the reaction path approximately follows \(\tilde x_A\).

We acknowledge support from U.S. National Science Foundation grants
DMR-9986948 (UCD, R.R.P.S.,S.K.P.) and PHY-9407194 (ITP Santa Barbara,
R.R.P.S. and D.L.C.), and U.S.
Department of Energy Grant DE-FG03-99ER45640 (D.L.C., R.G.E.). 
We are grateful for useful conversations with J.K. Barton,  G. Gruner, S. Isied,  J.B. Marston,
A.J. Millis,  P. Pincus, J. Rudnick, O. Schiemann, A. Stuc hebrukhov, and N. Wingreen. D.L.C. and
R.P.P.S. are grateful for the support and catalyzing influence of the workshop on ``Adaptive
Atoms in Biology, Chemistry, Environmental Science, and Physics'' at Los Alamos National Laboratory
sponsored by the Institute for Complex Adaptive Matter of the University of California.

\begin{figure}
\hspace{-1cm}
\psfig{file=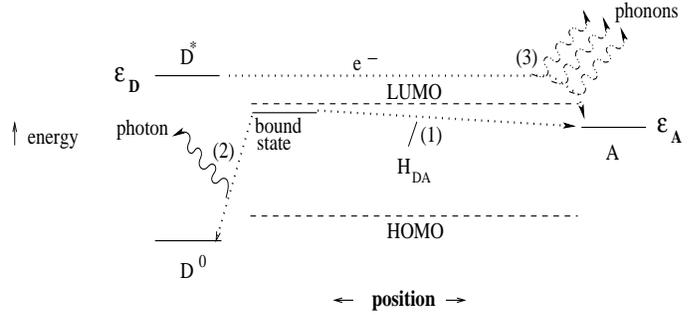,height=4.5cm,width=9.2cm}
\vspace*{0.05cm}
\caption{Dynamics of Kondo bound state (KS) mediated CT.
Photolysis leads to the excited state $D^0$ from which an electron is
transferred to the bridge leaving a remnant $D$ hole spin. 
For sufficiently rapid KS formation, KS mediated CT (1) is more rapid
and likely than luminescence (2) or phonon-radiative decay to the
acceptor (3).  }
\end{figure}    
         
\vspace{-1.0cm}

\begin{figure}
\psfig{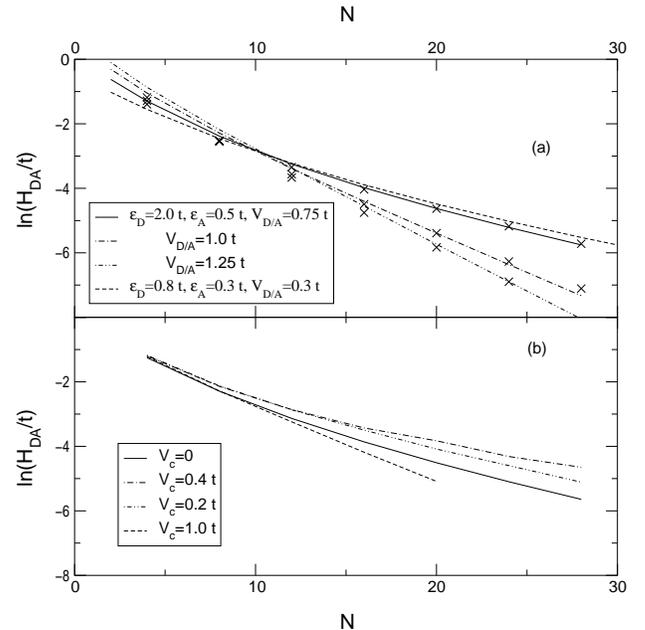}
\vspace*{0.05cm}
\caption{ Bridge length ($N$) dependence of electron tunnelling matrix 
element $H_{DA}$
(a) Dependence of $H_{DA}$  on bridge-donor coupling.
Lines:  variational method.
Symbols: DMRG approach.
Dotted line: 
$\epsilon_D$ and $V_D$ have been reduced so as to yield the same continuum
limit Kondo binding energy as the solid line.  
(B) Effect of Coulomb interaction $V_c$ between the bridge electron and
remnant D hole.}
\end{figure}

\vspace{-0.2cm}   
\begin{figure}
\hspace{-0.5cm}
\psfig{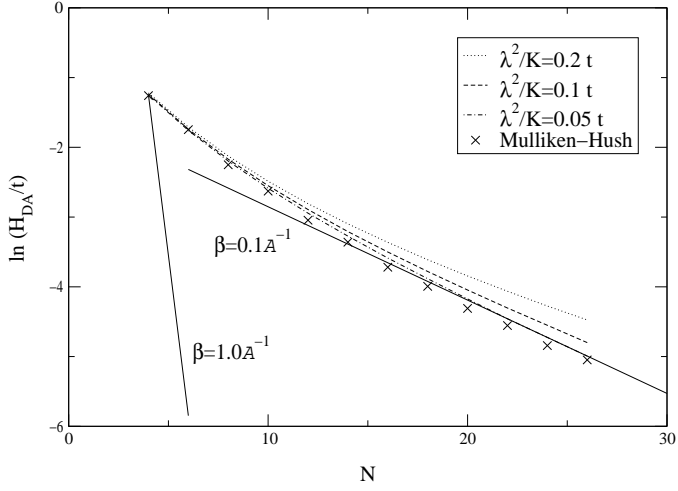}
\vspace*{0.05cm}
\caption{ Distance dependence of \(H_{DA}\): results from the splitting
of the Born-Oppenheimer surfaces 
are compared with the Mulliken-Hush algorithm as a function
of base pairs N for parameters \(\epsilon_D=2t\), \(\epsilon_A=0.5t\),
\(V=0.75t\). Thick solid lines:  exponential decay assuming $a=3.4\AA$}
\end{figure}


\begin{references}
\bibitem{vdnote} We estimate $V_D >t$ for 
metallointercalators\cite{barton,bartonrh}, 
$V_D\approx t$ for modified base pairs\cite{barzewail}, 
and $V_D\approx 0.5t$ for TET attached to PPV\cite{ratnernature}.
\end{references}
\end{document}